\documentclass[12pt,journal, twocolumn,romanappendices]{IEEEtran}
\usepackage{cite,calc,color}
\usepackage[cmex10]{amsmath} 
\usepackage{amsfonts,amsthm,amssymb,graphics,subfigure,epsfig,graphics,mathrsfs}
\usepackage[mathcal]{euscript} 
\usepackage[T1]{fontenc} 

\newtheorem{thm}{Theorem}

\newtheorem{rem}{Remark}

\theoremstyle{definition}
\newtheorem{defs}{Definition}

\newtheorem{lemma}{Lemma}

\DeclareMathOperator{\poly}{poly}

\newcommand\cX{\mathcal{X}}
\newcommand\cY{\mathcal{Y}}

\newcommand{\cC}{\mathcal{C}}

\newcommand{\cO}{{\mathcal{O}}}
\newcommand{\cE}{\mathcal{E}}

\newcommand{\cS}{\mathcal{S}}

\newcommand{\ck}{\text{\it k}}
\newcommand{\cK}{\text{\it K}}

\newcommand{\defeq}{\overset{\text{def}}{=}}
\newcommand{\bm}[1]{\mbox{\boldmath{$#1$}}}

\newcommand{\pr}{{\mathbb{P}}}
\newcommand{\ex}{{\mathbb{E}}}
\newcommand{\E}{\cE}
\newcommand{\openone}{\leavevmode\hbox{\small1\normalsize\kern-.33em1}}

\newcommand{\f}{\rho}
\newcommand{\ie}{{\it{i.e.,}}}

\begin{document}


\title{Sampling Constrained Asynchronous Communication: How to Sleep Efficiently}


\author{Venkat Chandar and Aslan Tchamkerten
\thanks{This work was supported in part by a grant and in part by a Chair of Excellence both
from the French National Research Agency (ANR-BSC and ANR-ACE, respectively). This work was presented in part at the $2015$ International Symposium on Information Theory and at the $2016$ Asilomar Conference on Signals, Systems, and Computers. }  \thanks{V.~Chandar
is with D. E. Shaw and Co., New York, NY 10036, USA. Email:
chandarvenkat@verizon.net.}  
\thanks{A.~Tchamkerten is with the
Department of Communications and Electronics, Telecom
ParisTech, 75634 Paris Cedex 13, France.
Email: aslan.tchamkerten@telecom-paristech.fr.}   }

\maketitle

{\begin{abstract} 

The minimum energy, and, more generally, the minimum cost, to transmit one bit of information has been recently derived for bursty
communication when information is available infrequently at random times at the
transmitter. Furthermore, it has been shown that even if the receiver
is constrained to sample only a fraction $\f\in (0,1]$ of
the channel outputs, there is no capacity penalty. That is, for any strictly positive sampling rate $\f$, the asynchronous capacity per unit
cost is the same as under full sampling, \ie, when $\f=1$. Moreover, there is no penalty in terms of decoding delay. 

The above results are asymptotic in nature, considering the limit as the number
$B$ of bits to be transmitted tends to infinity, while the sampling rate $\f$ remains fixed. A natural question is then whether the sampling rate $\f(B)$ can drop to zero without introducing
a capacity (or delay) penalty compared to full sampling. We answer this question affirmatively. The main result of this paper is an essentially tight characterization of the minimum sampling rate. We show that any sampling rate that grows at least as fast as
$\omega(1/B)$ is achievable, while any sampling rate smaller than $o(1/B)$ yields unreliable communication. The key ingredient in our improved achievability result is a new, multi-phase adaptive sampling scheme for locating transient changes, which we believe may be of independent interest for certain  change-point detection problems. 

\end{abstract}}


\normalsize
\section{Introduction}
\label{intro}


{ \IEEEPARstart{I}{n} many emerging technologies, communication is sparse and
asynchronous, but it is essential that when data is available, it is delivered 
to the destination as timely and reliably as possible. }   

 In \cite{chandar2013asynchronous} the authors characterized capacity per unit cost as a function of the level of asynchronism for the following model. There are $B$ bits of information that are made available to the 
transmitter at some random time $\nu$, and need to be
communicated to the receiver. The $B$ bits are encoded into a codeword of length 
$n$, and
transmitted over a memoryless channel using a sequence of symbols that have
costs associated with them. The rate $\bm{R}$ per unit cost is $B$ divided by the cost of the transmitted sequence.
Asynchronism is captured here by the fact that the random time $\nu$ is 
not known {\em a priori} to the receiver. However, both transmitter and receiver 
know that $\nu$ is distributed uniformly over a time horizon 
$\{1,2,\ldots,A\}$. At all times before and after the actual transmission, the 
receiver observes pure noise.

The goal of the receiver is to reliably decode the information bits by
sequentially observing the outputs of the channel.
A main result in \cite{chandar2013asynchronous} is a single-letter characterization of
the asynchronous capacity per unit cost
$\bm{C}(\beta),$ where $\beta=(\log A)/B$ denotes the 
{\emph{timing uncertainty per information bit}}. While this result holds for 
arbitrary discrete memoryless channels and
arbitrary input costs, the underlying model assumes that the receiver 
is always in the listening mode: every channel output is observed until the 
decoding instant. 

In \cite{6960771} it is shown that even if the receiver is constrained to observe at most a fraction $\rho\in (0,1]$ of the channel outputs the 
asynchronous capacity per unit cost $\bm{C}(\beta,\f)$ is {\emph{not}} impacted by a sparse 
output sampling, that is
$$\bm{C}(\beta,\f)=\bm{C}(\beta)$$
for any asynchronism level $\beta>0$ and sampling frequency $\rho\in (0,1]$. 
Moreover, the decoding delay is minimal: the elapsed time between when 
information is available sent and when it is decoded is asymptotically the same as under full 
sampling. This result uses the possibility for the receiver to sample 
adaptively: the next sample can be chosen
as a function of past observed samples. In fact, under non-adaptive
sampling, it is still possible to achieve the full sampling asynchronous 
capacity per unit cost, but the decoding delay gets multiplied by a factor 
$1/\rho$. Therefore, adaptive sampling strategies are of particular interest in 
the very sparse regime.

The results of \cite{6960771} provide an achievability scheme when the sampling
frequency $\f$ is a strictly positive constant. This suggests the question whether $\f=\f(B)$ can tend to zero as $B$ tends to infinity while still 
incurring no capacity or delay penalty. The main result of this paper resolves
this question. We introduce a novel, multi-phase adaptive sampling algorithm for message detection, 
and use it to prove an essentially tight asymptotic characterization of 
the minimum sampling rate needed in order to communicate as efficiently 
as under full sampling. Informally, we exhibit a communication
scheme utilizing this multi-phase sampling method at the receiver 
that asymptotically achieves vanishing probability of error and possesses
the following properties:
\begin{itemize}
\item[1.] The scheme achieves the capacity per unit cost under 
full sampling, that is, there is no rate penalty even though the sampling rate tends to zero;
\item[2.] The receiver detects the codeword with minimal delay;
\item[3.] The receiver detects changes with minimal sampling rate, in the 
sense that any 
scheme that achieves the same order of delay but operates at a lower sampling rate will 
completely miss the codeword transmission period, regardless of 
false-alarm probability. The sampling rate 
converges to $0$ in the limit of large $B$, and our main result characterizes
the best possible rate of convergence.
\end{itemize}
In other words, our communication scheme achieves essentially the minimal sampling
rate possible, and incurs no delay or capacity penalty relative to full
sampling. A formal statement of the main result is given in Section \ref{scta}.
\subsection*{Related works}
The above sparse communication model was first introduced in \cite{chandar2008optimal,tchamkerten2009communication}. These works characterize the {\emph{synchronization threshold}}, \ie the largest level of asynchronism under which it is still possible to communicate reliably.  In \cite{tchamkerten2009communication, 6352910} capacity is defined as the message length divided by the mean elapsed time between when information is available and when it is decoded. For this definition, capacity upper and lower bounds are established and shown to be tight for certain channels.  In \cite{6352910} it is also shown that so called training-based schemes, where synchronization and information transmission are performed separately,  need not be optimal in particular in the high rate regime. 
In \cite{chandar2013asynchronous} capacity is defined with respect to codeword length and is characterized as a function of the level of asynchronism. For the same setup  Polyanskiy in \cite{polyanskiy2013asynchronous} investigated the finite length regime and showed that in certain cases dispersion is unaffected by asynchronism even when $\beta>0$. 

In \cite{wang2010distinguishing,6848789} the authors investigated the slotted version of the problem ({\it{i.e.}}, the decoder is revealed $\nu\,\text{mod}\, n$) and established error exponent tradeoffs between between decoding error, false-alarm, and miss-detection.

In \cite{chandar2013asynchronous,shahi2016capacity} the above bursty communication setup is investigated in a random access configuration and tradeoffs between communication rate and number of users are derived as a function of the level of asynchronism. Finally,  in \cite{shomorony2012bounds} a diamond network is considered and the authors provided bounds on the minimum energy needed to convey one bit across the network.

\subsection*{Paper organization}
This paper is organized as follows. 
In Section \ref{scta}, we recall the asynchronous communication model and related prior results. Then, we state
our main result, Theorem~\ref{thm}, which is a stronger version of the results in 
\cite{6960771}.  Section~\ref{moper} states 
auxiliary results, Theorems~\ref{unow} and \ref{unowaa},  characterizing the performance of our multi-phase sampling 
algorithm. In Section~\ref{analysis} we first prove Theorems~\ref{unow} and \ref{unowaa}, then prove Theorem~\ref{thm}. The achievability part of Theorem~\ref{thm} uses the multi-phase sampling algorithm for message detection at the receiver, and the converse is essentially an immediate consequence of the converse of Theorem~\ref{unowaa}.

\section{Main result: the sampling rate required in asynchronous communication}\label{scta}
Our main result, Theorem~\ref{thm} below, is a strengthening of the results of \cite{6960771}. We recall the model and results (Theorems~\ref{fullsampling} and \ref{delayp}) of that paper below to keep the paper self-contained.

Communication is discrete-time and carried over a discrete memoryless
channel characterized by its finite input and output alphabets $$\cX\quad \text{and}\quad \cY\,,$$ respectively, and transition probability matrix $$Q(y|x),$$ for
all $y\in \cY$ and $x\in \cX$.  Without loss of generality, we assume
that for all $y\in \cY$ there is some $x\in \cX$ for which $Q(y|x)>0$.

Given $B\geq 1 $ information bits to be transmitted, a codebook ${\cal{C}}$ consists of
$$M= 2^B$$ codewords of length $n\geq 1$ composed of symbols from ${\cal{X}}$. 

A randomly and uniformly chosen message $m$ is available at the transmitter at a random time $\nu$,
independent of $m$, and uniformly distributed over
$\{1,\ldots,A_B\}$, where the integer $$A=2^{\beta B}$$ characterizes the {\emph{asynchronism
level}} between the transmitter and the receiver, and
where the constant $$\beta\geq 0$$ denotes the
{\emph{timing uncertainty per information bit}. While $\nu $ is unknown to the receiver, $A $ is known by both the transmitter and the receiver.

We consider one-shot communication, \ie only one message arrives
over the period $\{1,2, \ldots, A \}\,.$  If $A =1$, the
channel is said to be synchronous.

Given $\nu $ and $m$, the transmitter chooses a time
$\sigma (\nu ,m)$ to start sending codeword $c^n(m)\in
\cC$ assigned to message $m$. Transmission cannot
start before the message arrives or after the end of
the uncertainty window, hence  $\sigma (\nu ,m)$ must
satisfy $$\nu  \leq \sigma  (\nu ,m)\leq A \quad
\text{almost surely.}$$ In the rest of the paper, we
suppress the arguments $\nu $ and $m$ of $\sigma $ when
these arguments are clear from context. 

Before and after the codeword transmission, \ie before time $\sigma $ and after time $\sigma +n-1$, the receiver
observes ``pure noise.'' Specifically, conditioned on $\nu $ and on the message to be conveyed $m$, the receiver
observes independent channel outputs $$Y_1,Y_2,\ldots,Y_{A +n-1}$$ distributed as
follows. For $$1\leq t\leq \sigma -1$$ or $$\sigma +n\leq t\leq A +n-1\,,$$
the $Y_t$'s are ``pure noise'' symbols, \ie$$ Y_t\sim Q(\cdot|\star)$$ where $\star$ represents the ``idle'' symbol.  For $
\sigma  \leq t\leq  \sigma +n-1$
$$Y_t\sim Q(\cdot|{c_{t-\sigma +1}(m)})$$ where $c_{i}(m)$ denotes the $i$\/th symbol
of the codeword $c^n(m)$.

Decoding involves three components: 
\begin{itemize}
\item
a sampling strategy,
\item
 a stopping (decoding) time defined on the sampled process, 
 \item
 a decoding function defined on the stopped sampled process. 
\end{itemize}
A sampling
strategy consists of ``sampling times'' which are
defined as an ordered collection of random time indices
$${\cS}=\{(S_1,\ldots,S_\ell)\subseteq
\{1,\ldots,A +n-1\}:S_i< S_j,i<j\}$$  where $S_j$ is
interpreted as the $j$\/th sampling time. The sampling
strategy is either non-adaptive or adaptive. It is non-adaptive when the
sampling times in ${\cS}$ are independent
of $Y_1^{A +n-1}$. The strategy is adaptive when the sampling times are functions of past observations. This means that $S_1$ is an arbitrary value in $\{1,\ldots, A +n-1\}$, possibly random
but independent of $Y_1^{A +n-1}$, and for $j\geq 2$ $$S_j=g_j(\{Y_{S_i}\}_{i<j})  $$
for some (possibly randomized) function $$g_j:
\cY^{j-1}\to \{S_{j-1}+1,\ldots,A +n-1\}\, .$$

Given a sampling strategy, the receiver decodes by means of a sequential test
$(\tau,\phi_\tau)$ where $\tau$ denotes a stopping (decision) time with respect to the sampled output process\footnote{Recall that a (deterministic or
randomized) stopping time $\tau$ with respect to a sequence of random variables
$Y_1,Y_2,\ldots$ is a positive, integer-valued, random variable such that the
event $\{\tau=t\}$, conditioned on the
realization of $Y_1,Y_2,\ldots,Y_t$, is
independent of the realization of
$Y_{t+1},Y_{t+2},\ldots$ for all $t\geq 1$.}
 $$Y_{S_1},Y_{S_2},\ldots$$ and where $\phi_\tau$ denotes a decoding function based on the stopped sampled output process.  Let
 \begin{align}\label{st}
\cS^{ t}\defeq \{S_i\in \cS:S_i\leq t\}.
\end{align}
 denote the set of sampling times taken up to time $t$ and let
\begin{align}\label{sto}
\cO^t\defeq \{Y_{S_i}: S_i\in \cS^{t}\}
\end{align}
denote the corresponding set of channel outputs.
The decoding function $\phi_\tau$ is a map \begin{align*}
 \phi_\tau: \cY^{|\cO_\tau|}&\to \{1,2,\ldots,M\}\\
 \cO^\tau&\mapsto \phi_\tau( \cO^\tau).
\end{align*}

A code $(\cC,(\cS,\tau,\phi_\tau))$ is defined as a codebook and a decoder composed of 
a sampling strategy, a decision time, and a decoding function. Throughout the paper, whenever clear from context, we
often refer to a code using the codebook symbol $\cC$
only, leaving out an explicit reference to the decoder.

Note that a pair $(\cS,\tau)$ allows only to do message detection but does not provide a message estimate.  Such a restricted decoder will later (Section~\ref{moper})  be referred simply as a ``detector.''

\begin{defs}[Error probability]
The maximum (over messages) decoding error probability
of a code $\cC$ is defined as
\begin{align}\label{maxerror}
\max_m\pr_m({\EuScript{E}}_m|\cC),
\end{align}
where 
$$\pr_m({\EuScript{E}}_m|\cC) \defeq \frac{1}{A }\sum_{t=1}^{A }
\pr_{m,t} ({\EuScript{E}}_m|\cC),$$
where the subscripts ``$m,t$'' denote conditioning on the
event that message $m$ arrives at
time $\nu =t$, and where $\EuScript{E}_m$ denotes the
error event that the decoded message does not
correspond to $m$, \ie
\begin{align}\label{erevent}
\E_m\defeq\{\phi_\tau(\cO^\tau)\ne m\}\,.
\end{align}
\end{defs}

\begin{defs}[Cost of a code]\label{costc}
The (maximum) cost of a code ${\cal C}$ with respect to a cost function $\ck: \cX\to [0,\infty]$ is defined as
$$ \cK({\cal C})\defeq \max_m \sum_{i=1}^n \ck(c_i(m)).$$
\end{defs}

\begin{defs}[Sampling frequency of a code]\label{costc}
Given $\varepsilon>0$, the sampling frequency of a code $\cC$, denoted by
$\rho ({\cal{C}},\varepsilon)$, is the relative number of channel outputs that are observed until a message is
declared. Specifically, it is defined as the minimum $r\geq 0$ such that
 $$\min_m\pr_m(|\cS_\tau|/{\tau}\leq
 r)\geq1-\varepsilon\,.$$
\end{defs}

\begin{defs}[Delay of a code]\label{def:delaiscode}
Given $\varepsilon>0$, the (maximum) delay of a code ${\cal
C}$, denoted by $d({\cal
C}, \varepsilon)$, is defined as the minimum integer $l$ such that
$$\min_m\pr_m(\tau-\nu \leq  l-1) \geq
1-\varepsilon\,.$$
\end{defs}

We now define capacity per unit cost under the constraint that the
receiver has  access to a limited number of channel outputs:

\begin{defs}[Asynchronous capacity per unit cost under sampling constraint]
\label{def:cap}
Given $\beta\geq 0$ and a non-increasing sequence of numbers $\{\rho_B \}$, with $0\leq \rho_B \leq 1$,
rate per unit cost $\bm{R}$ is said to be achievable  if
 there exists a sequence of codes $\{{\cal C}_B\}$
 and a sequence of positive numbers
 $\varepsilon_B$ with $\varepsilon_B\overset{B\to \infty}{\longrightarrow}0$
 such that for all $B$ large enough
 \begin{enumerate}
\item $\cC_B$ operates at timing uncertainty per information bit $\beta$;
\item the maximum error probability $\pr(\cE|{\cal{C}}_B)$ is at most
$\varepsilon_B$;
\item the rate per unit cost $$\frac{B}{\cK({\cal C}_B)}$$ is at least $
\bm{R}-\varepsilon_B$;
\item the sampling frequency satisfies
$$\rho({\cal{C}}_B,\varepsilon_B)\leq \rho_B; $$
\item the delay satisfies\footnote{Throughout the paper logarithms are always intended to be to the base
$2$.}
 $$ \frac{1}{B}\log(d({\cal
C}_B,\varepsilon_B)) \leq \varepsilon_B\,.$$
\end{enumerate}
Given $\beta$ and $\{\rho_B\}$, the asynchronous capacity per unit cost, denoted by $\bm{C}(\beta,\{\rho_B\})$,
is the supremum of achievable rates per unit cost. 
\end{defs}

Two comments are in order. First note that samples occurring  after time $\tau$ play no role in our performance metrics since error probability, delay, and sampling rate are are all functions of $\cO^{\tau}$ (defined in \eqref{sto}). Hence, without loss of generality, for the rest of the paper we assume that the last sample is taken at time $\tau$, \ie that the sampled process is truncated at time $\tau$.
 The truncated sampled process is thus given by the collection of sampling times $\cS^{\tau}$ (defined in \eqref{st}). 
In particular, we have (almost surely)
\begin{align}\label{trunsam2}
\cS^1\subseteq \cS^2 \subseteq \cdots \subseteq \cS^{\tau}= \cS^{\tau+1}=\cdots =\cS^{A_B+n-1}.\end{align}

The second comment concerns the delay constraint $4)$. The delay constraint is meant to capture the fact that the receiver is able to locate $\nu_B$ with high accuracy. More precisely, with high
probability, $\tau_B$ should be at most sub-exponentially larger than $\nu_B$. This already represents a decent level of accuracy, given that $\nu_B$ itself is uniform over an exponentially large interval. However, allowing a sub-exponential delay still seems like a very loose constraint. As Theorem \ref{thm} claims, however, we can achieve much greater accuracy. Specifically, if a sampling rate is achievable, it can be achieved with delay linear in $B$, and if a sampling rate cannot be achieved with linear delay, it cannot be achieved even if we allow a sub-exponential delay.

\noindent{\it Notational conventions:}
We shall use $d_B$ and $\rho_B$ instead of $d(\cC_B,\varepsilon_B)$ and $\rho(\cC_B,\varepsilon_B)$, respectively, leaving out any explicit reference to $\cC_B$ and the sequence of non-negative numbers $\{\varepsilon_B\}$, which we assume satisfies $\varepsilon_B\to 0$. Under full sampling, \ie when $\rho_B=1$ for all $B$, capacity is simply denoted by $\bm{C}(\beta)$, and when the sampling rate is constant, \ie when $\rho_B=\rho\leq 1$ for all $B$, capacity is denoted by $\bm{C}(\beta,\rho)$.

The main, previously known, results regarding capacity for this asynchronous communication model 
are the following.
First, capacity per unit cost under full sampling is given by the following 
theorem:
\begin{thm}[Full sampling, Theorem 1~\cite{6397617} \label{fullsampling}]
For any $\beta\geq 0$
\begin{align}\label{capexpr}
 \bm{C}(\beta)=\max_X \min\left\{\frac{I(X;Y)}{\ex[\ck(X)]},
\frac{I(X;Y ) + D(Y||Y_\star)}{\ex[\ck(X)](1 + \beta)}\right\}
\end{align}
where $\max_X$ denotes maximization with respect to the channel input distribution $P_X$, where 
$(X,Y) \sim P_X(\cdot) Q(\cdot|\cdot)$,  where $Y_\star$ denotes the random output of the channel when the
idle symbol $\star$ is transmitted (\ie $Y_\star\sim Q(\cdot|\star)$), where
$I(X;Y)$ denotes the mutual information between $X$ and $Y$, and where $D(Y||Y_\star)$ denotes the divergence
between the distributions of $Y$ and~$Y_\star$. \hfill \QED
\end{thm}
Theorem~\ref{fullsampling} characterizes capacity per unit cost under full 
output sampling, and over codes whose delay grow sub-exponentially with $B$. As 
it turns out, the full sampling capacity per unit cost can also be achieved with linear delay
and sparse output sampling.

Define\footnote{Throughout the paper we use the standard ``big-O'' Landau notation to characterize 
growth rates  (see, e.g., \cite[Chapter~3]{CLRS}). These growth rates, {\it{e.g.}}, $\Theta(B)$ or $o(B)$, are intended in the limit $B\to \infty$, unless stated otherwise. }
\begin{align}\label{ennstar}n^*_B(\beta, \bm{R})\defeq  \frac{B}{\bm{R}\max \{ \ex [k(X)]: {X}\in {\cal{P}}(\bm{R})\}} =\Theta(B)
\end{align}
where ${\cal{P}}(\bm{R})$ is defined as the set \begin{align}
\label{probset}
\left\{X:\min\left\{\frac{I(X;Y)}{\ex[\ck(X)]} ,
\frac{I(X;Y ) + D(Y||Y_\star)}{\ex[\ck(X)](1 + \beta)}\right\}\geq \bm{R}\right\}.
\end{align}
The quantity $n^*_B(\beta,\bm{R})$ quantifies the minimum detection delay as a 
function of the asynchronism level and rate per unit cost, under full sampling:

\begin{thm}[Minimum delay, constant sampling rate, Theorem 3 \cite{6960771}]\label{delayp}
Fix $\beta\geq 0$,  $\bm{R}\in (0,\bm{C}(\beta)]$, and $\f\in (0,1]$. For any codes $\{{\cal{C}}_B\}$ that achieve rate per unit cost $\bm{R}$ at timing uncertainty $\beta$, and operating at constant sampling rate $0<\f_B=\rho$, we have
$$\liminf_{B\to \infty}\frac{d_B}{n^*_B(\beta,\bm{R})}\geq 1.$$
 Furthermore,  there exist codes $\{{\cal{C}}_B\}$ that achieve rate $\bm{R}$ with (a) timing uncertainty $\beta$, (b) sampling rate $\f_B=\f$, and (c) delay 
$$\limsup_{B\to \infty}\frac{d_B}{n^*_B(\beta,\bm{R})}\leq 1.$$
\end{thm}

Theorem \ref{delayp} says that the minimum delay achieved by rate $\bm{R}\in (0,\bm{C}(\beta)]$ codes is $n^*_B(\beta,\bm{R})$ for any constant sampling rate $\rho\in (0,1]$.
This naturally suggests the question ``What is the minimum sampling rate of codes that achieve rate $\bm{R}$ and minimum delay $n^*_B(\beta, \bm{R})$?'' Our main result is the following theorem, which states that the minimum sampling rate essentially decreases as $1/B$:

\begin{thm}[Minimum delay, minimum sampling rate] \label{thm}

Consider a sequence of codes $\{\cC_B\}$ that operate under timing uncertainty per information bit $\beta> 0$.  If \begin{align}
\rho_Bd_B=o(1),\label{nece}
\end{align} the receiver does not even sample a single component of the sent codeword with probability tending to one. Hence, the average error probability tends to one whenever  $\bm{R}>0$, $d_B=O(B)$, and $\f_B=o(1/B)$.

Moreover, for any $\bm{R}\in (0,\bm{C}(\beta)]$ and any sequence of sampling rates satisfying $\f_B=\omega(1/B)$,  there exist codes $\{\cC_B\}$ that achieve rate $\bm{R}$ at (a) timing uncertainty $\beta$, (b) sampling rate $\f_B$, and (c) delay $$\limsup_{B\to \infty}\frac{d_B}{n^*_B(\beta,\bm{R})}\leq 1.$$

\end{thm}
If $\bm{R}>0$, the minimum delay $n^*_B(\beta,\bm{R})$ is $O(B)$ by Theorem~\ref{delayp} and \eqref{ennstar}, so Theorem~\ref{thm} gives an essentially tight characterization of the minimum sampling rate; a necessary condition for achieving the minimum delay is that $\f_B$ be at least $\Omega(1/B)$, and any $\f_B=\omega(1/B)$ is sufficient. 

That sampling rates of order $o(1/d_B)$ are not achievable is certainly intuitively plausible and even essentially trivial to prove when restricted to non-adaptive sampling. To see this note that by the definition of delay, with high probability decoding happens no later than instant $\nu+d_B$.  Therefore, without essential loss of generality, we may assume that information is being transmitted only within period $\{\nu, \nu+1, \ldots,\nu+d_B\}$. Hence, if sampling is non-adaptive and its rate is of order $o(1/d_B)$ then with high probability (over $\nu$) information transmission will occur during one unsampled period of duration $d_B$. This in turn implies a high error probability. The main contribution in the converse argument is that it also handles adaptive sampling.

Achievability rests on a new multi-phase procedure to efficiently detect the sent message. This detector, whose performance is the focus of Section~\ref{moper}, is a much more fine grained procedure than the one used to establish Theorem~\ref{delayp}. To establish achievability of Theorem~\ref{delayp}, a two-mode  detector is considered, consisting of a baseline mode operating at low sampling rate, and a high rate mode.  The detector starts in the baseline mode and, if past observed samples suggest the presence of a change in distribution, the detector changes to the high rate mode which acts as a confirmation phase. At the end of the confirmation phase the detector either decides to stop, or decides to reverse to the baseline mode in case the change is unconfirmed. 

The detector proposed in this paper (see Section~\ref{moper} for the setup and Section~\ref{ilcuore} for the description of the procedure) has multiple confirmation phases, each operating at a higher sampling rate than the previous phase. Whenever a confirmation phase is passed, the detector switches to the next confirmation phase. As soon as a change is unconfirmed, the procedure is aborted and the detector returns to the low rate baseline mode. The detector only stops if the change is confirmed by all confirmation phases. Having multiple confirmation phases instead of just one, as for Theorem~\ref{delayp}, is key to reducing the rate from a  constant to essentially $1/B$, as it allows us to aggressively reject false-alarms whithout impacting the ability to detect the message.    


\section{Sampling constrained transient change-detection}\label{moper}
This section focuses on one key aspect of asynchronous communication, namely, that we need to quickly detect the presence of a message with a sampling constrained detector. As there is only one possible message, the problem amounts to a pure (transient) change-point detection problem. Related results are stated in Theorems~\ref{unow} and~\ref{unowaa}. These results and their proofs are the key ingredients for proving Theorem~\ref{thm}. 

\subsection{Model}\label{siz}
The transient change-detection setup we consider in this section is essentially a simpler 
version of the asynchronous communication problem stated in Section~\ref{scta}.
Specifically, rather than having a codebook of $2^B$ messages, we consider a
binary hypothesis testing version of the problem. There is a single codeword,
so no information is being conveyed, and our goal is simply to detect when the 
codeword was transmitted.

Proceeding more formally, let $P_0$ and $P_1$ be distributions defined over some finite alphabet $\cal{Y}$ and with finite divergence
$$D(P_1||P_0)\defeq \sum_y P_1(y)\log [P_1(y)/P_0(y)].$$ 

There is no parameter $B$ in our problem, but in analogy with Section~\ref{scta}, let $n$ denote the length of the transient change. Let $\nu$ be uniformly distributed over $$\{1,2,\ldots,A=2^{\alpha n}\}.$$ where the integer $A$ denotes the {\emph{uncertainty level}} and where $\alpha$ the corresponding {\emph{uncertainty exponent}}, respectively. 
 
 Given $P_0$ and $P_1$, process $\{Y_t\}$ is defined similarly as in Section~\ref{scta}. Conditioned on the value of $\nu$, the $Y_t$'s are i.i.d. according to    $ P_0$ for  $$1\leq t<\nu$$ or $$\nu_n+n\leq t \leq A+n-1$$ and i.i.d. according to  $ P_1$ for $\nu\leq t\leq \nu+n-1$. Process $\{Y_t\}$ is thus i.i.d.~$P_0$ except for a brief period of duration $n$ where it is i.i.d. $P_1$.

Sampling strategies are defined as in Section~\ref{scta}, but since we now
only have a single message, we formally define the relevant performance metrics
below.

\begin{defs}[False-alarm probability] For a given detector $(\cS,\tau)$
the probability of false-alarm is defined as
$$\pr(\tau <\nu)=\pr_0(\tau<\nu)$$
where $\pr_0$ denotes the joint distribution over $\tau$ and $\nu$ when the observations are
drawn from the $P_0$-product distribution. In other words, the false-alarm probability is the probability
that the detector stops before the transient change has started.
\end{defs}

\begin{defs}[Detection delay]\label{def:delaiscode0}
For a given detector $(\cS,\tau)$ and $\varepsilon>0$, the delay, denoted by $d((\cS,\tau), \varepsilon)$, is defined as the minimum $l\geq 0$ such that
$$\pr(\tau-\nu \leq  l-1) \geq
1-\varepsilon\,.$$
\end{defs}
\noindent\emph{Remark:} The reader might wonder why we chose the above definition of delay, as opposed to, for example, measuring delay by $\ex[\max(0, \tau-\nu)]$. The above definition corresponds to capturing the ``typical'' delay, without incurring a large penalty in the tail event where $\tau$ is much larger than $\nu$, say because we missed the transient change completely. We are able to characterize optimal performance tightly with the above definition, but expected delay would also be of interest, and an 
analysis of the optimal performance under this metric is an open problem for future research.

\begin{defs}[Sampling rate]\label{costc}
For a given detector $(\cS,\tau)$ and $\varepsilon>0$, the sampling rate, denoted by
$\rho ((\cS,\tau),\varepsilon)$, is defined as the minimum $r\geq 0$ such that
 $$\pr(|\cS^\tau|/{\tau}\leq
 r)\geq1-\varepsilon.$$
\end{defs}

Achievable sampling rates are defined analogously to Section~\ref{scta}, but we include a formal definition for completeness.
\begin{defs}[Achievable sampling rate]
\label{def:capdetc}
Fix $\alpha\geq 0$, and fix a sequence of non-increasing values $\{\f_n\}$ with $0\leq \f_n\leq 1$. 
Sampling rates $\{\f_n\}$ are said to be achievable at uncertainty exponent $\alpha$ if there exists a sequence of detectors
$\{(\cS_n,\tau_n)\}$ such that for all $n$ large enough 
\begin{enumerate}
\item
$(\cS_n,\tau_n)$ operates under
uncertainty level $A_n=2^{\alpha n}$,
\item
the false-alarm probability $\pr(\tau_n<\nu_n)$ is at most $\varepsilon_n$,
\item
 the sampling rate satisfies $\rho((\cS_n,\tau_n),\varepsilon_n)\leq \f_n$, 
 \item
 the delay satisfies $$ \frac{1}{n}\log(d((\cS_n,\tau_n),\varepsilon_n)) \leq \varepsilon_n$$
\end{enumerate}
for some sequence of non-negative numbers  $\{\varepsilon_n\}$ such that~$\varepsilon_n\overset{n\to \infty}{\longrightarrow}0$.
\end{defs}

\noindent{\it Notational conventions:}
We shall use $d_n$ and $\rho_n$ instead of $d((\cS_n,\tau_n),\varepsilon_n)$ and $\rho((\cS_n,\tau_n),\varepsilon_n)$, respectively, leaving out any explicit reference to the detectors and the sequence of non-negative numbers $\{\varepsilon_n\}$, which we assume satisfies $\varepsilon_n\to 0$. 
\subsection{Results}
Define \begin{align}\label{nstaar}n^*(\alpha)\defeq \frac{n\alpha}{D(P_1||P_0)}=\Theta(n).\end{align}
\begin{thm}[Detection, full sampling]\label{unow}
Under full sampling ($\f_n=1$):
\begin{enumerate}
\item
 the supremum of the set of achievable uncertainty exponents is $D(P_1||P_0)$; 
 \item  any detector that achieves uncertainty exponent $\alpha\in (0,D(P_1||P_0))$ has a delay that satisfies
 $$\liminf_{n\to \infty}\frac{d_n}{n^*(\alpha)}\geq 1;$$
 \item any uncertainty exponent $\alpha\in (0,D(P_1||P_0))$ is achievable with delay satisfying
  $$\limsup_{n\to \infty}\frac{d_n}{n^*(\alpha)}\leq 1.$$
 \end{enumerate}
\end{thm}
Hence, the shortest detectable\footnote{By detectable we mean with vanishing false-alarm probability and subexponential delay.} change  is of size \begin{align}\label{minle}
n_{\min}(A_n)=\frac{\log A_n}{D(P_1||P_0)}(1\pm o(1))
\end{align}by Claim 1) of Theorem~\ref{unow}, assuming $A_n\gg 1$. In this regime, change duration and minimum detection delay are essentially the same by Claims 2)-3) and \eqref{nstaar}, \ie 
$$n^*(\alpha=(\log A_n)/n_{\min}(A_n))=n_{\min}(A_n)(1\pm o(1))$$
whereas in general minimum detection delay could be smaller than change duration.

The next theorem says that the minimum sampling rate needed to achieve the same detection delay as under full sampling decreases essentially as $1/n$. Moreover, any detector that tries to operate below this sampling limit will have a huge delay.
\begin{thm}[Detection,  sparse sampling]\label{unowaa}
Fix $\alpha\in (0,D(P_1||P_0))$. Any sampling rate $$\f_n=\omega(1/n)$$  is achievable with delay satisfying $$\limsup_{n\to \infty}\frac{d_n}{n^*(\alpha)}\leq 1.$$
Conversely, if
 $$\rho_n=o(1/n)$$
the detector samples only from distribution $P_0$ (\ie it completely misses the change) with probability tending to one. This implies that the delay is $\Theta(A_n=2^{\alpha n})$ whenever the probability of false-alarm tends to zero.\end{thm}

\section{Proofs}\label{analysis}

\subsection*{Typicality convention}
A length $q\geq 1$ sequence $v^q$ over ${\cal{V}}^q$ is said to be typical with respect to some distribution ${P}$ over  ${\cal{V}}$ if\footnote{$||\cdot ||$ refers to the $L_1$-norm.}
$$||  \hat{P}_{v^q}-{P}|| \leq q^{-1/3} $$ where $\hat{P}_{v^q}$ denotes the empirical distribution (or type) of $v^q$. 
  
Typical sets have large probability. Quantitatively, a simple consequence of Chebyshev's inequality
is that
\begin{align}\label{typset2}
{P}^q(||\hat{P}_{V^q}-{P}||\leq q^{-1/3} ) = 1 - O\left(q^{-1/3}\right) \: (q\to \infty)
\end{align}
where ${P}^q$ denotes the $q$-fold product distribution of ${P}$.
Also, for any distribution $\tilde{P}$ over $\cal{V}$
we have 
\begin{align}\label{mismatch}{P}^q(||\hat{P}_{V^q}-\tilde{P}||\leq q^{-1/3} ) \leq 2^{-q (D(  \tilde{P}|| {P} )-o(1))}.\end{align}

\subsection*{About rounding}
Throughout computations, we ignore issues related to the rounding of non-integer quantities, as they play no role asymptotically.

\subsection{Proof of Theorem~\ref{unow}}
The proof of Theorem~~\ref{unow} is essentially a Corollary of \cite[Theorem]{chandar2008optimal}. We sketch the main arguments.

\subsubsection{}
To establish achievability of $D(P_1||P_0)$ one uses the same sequential typicality detection procedure as in the achievability of \cite[Theorem]{chandar2008optimal}. For the converse argument, we use similar arguments as for the converse of \cite[Theorem]{chandar2008optimal}. For this latter setting, achieving $\alpha$ means that we can drive the probability of the event $\{\tau_n\ne \nu_n+n-1\}$ to zero. Although this performance metric differs from ours---vanishing probability of false-alarm and sub-exponential delay---a closer look at the converse argument of  \cite[Theorem]{chandar2008optimal} reveals that if $\alpha>D(P_1||P_0)$ there are exponentially many sequences of length $n$ that are ``typical'' with respect to the posterior distribution. This, in turn, implies that either the probability of false-alarm is bounded away from zero, or the delay is exponential.
\subsubsection{}

Consider stopping times
$\{\tau_n\}$ that achieve delay $\{d_n\}$, and vanishing false-alarm probability  (recall the notational conventions for $d_n$ at the end of Section~\ref{siz}). 
We define the
``effective process'' $\{\tilde{Y}_i\}$ as the process whose change has duration $\min\{d_n,n\}$ (instead of $n$).

\noindent{\emph{Effective output process:}} The effective process $\{\tilde{Y}_i\}$ is defined as follows. Random variable $\tilde{Y}_i$ is equal to $Y_i$ for any index $i$ such that
$$ 1\leq i\leq \nu_n+\min\{d_n,n\}-1$$
and 
$$\{\tilde{Y}_i: \nu_n+\min\{d_n,n\}\leq i\leq A_n+n-1\}
$$
is an i.i.d. $P_0$ process independent of $\{{Y}_i\}$.
Hence, the effective process differs from the true process over the period $\{1,2,\ldots, \tau_n\}$ only when $\{\tau_n\geq \nu_n+d_n\}$ with $d_n< n$. 

\noindent{\emph{Genie aided statistician:}}
A genie aided statistician observes the entire effective process (of duration $A_n+n-1$) and is informed that the change
occurred over one of \begin{align}\label{rr0}
r_n\defeq \left\lfloor \frac{A_n+n-1-(\nu_n \mbox{
mod } d_n)}{d_n}\right\rfloor
\end{align}
 consecutive (disjoint) blocks of
duration $d_n$. 
The genie aided statistician produces a time interval of size $d_n$ which corresponds to an estimate of the change in distribution and is declared to be correct only if this interval corresponds to the change in distribution.

Observe that since $\tau_n$ achieves false-alarm probability $\varepsilon_n$ and delay ${d_n}$ on the true process $\{{Y}_i\}$, the genie aided statistician achieves error probability at most $2\varepsilon_n$. The extra $\varepsilon_n$ comes from the fact $\tau_n$ stops after time $\nu_n+d_n-1$ (on $\{{Y}_i\}$) with probability at most $\varepsilon_n$. Therefore, with probability at most $\varepsilon_n$ the genie aided statistician observes a process that may differ from the true process.

By using the same arguments as for the converse of  \cite[Theorem]{chandar2008optimal}, but on the process $\{\tilde{Y}_i\}$ parsed into consecutive slots of size $d_n$, we can conclude that if $$\liminf_{n\to \infty}\frac{d_n}{n^*(\alpha)}<1$$ then the error probability of the genie aided decoder tends to one.

\subsubsection{}
 To establish achievability apply the same sequential typicality test as in the achievability part of \cite[Theorem]{chandar2008optimal}. \hfill $\small\blacksquare$

\subsection{Proof of Theorem~\ref{unowaa}: Converse}\label{cth2}
As alluded to earlier (see discussion after Theorem~\ref{thm}), it is essentially trivial to prove that sampling rates of order $o(1/n)$ are not achievable when we restrict to non-adaptive sampling, that is when all sampling times are independent of $\{Y_t\}$.  The main contribution of the converse, and the reason why it is somewhat convoluted, is that it handles adaptive sampling as well.

Consider a sequence of detectors $\{(\cS_n,\tau_n)\}$ that achieves, for some false-alarm probability $\varepsilon_n\to 0$,  sampling rate $\{\f_n\}$ and
communication delay $d_n$ (recall the notational conventions for $d_n$ and $\rho_n$ at the end of Section~\ref{siz}). 

We show first that if 
\begin{align}\label{rdel2}
\rho_n=o(1/n)
\end{align}
then any detector, irrespective of delay, will take only $P_0$-generated samples with probability asymptotically tending to one. This, in turn, will imply that
the delay is exponential, since by assumption the false-alarm probability vanishes.

In the sequel, we use $\pr(\cdot)$ to denote the (unconditional) joint distribution of the output process $Y_1,Y_2,\ldots$ and $\nu$, and we use $\pr_0(\cdot)$ to denote the distribution of the output process $Y_1,Y_2,\ldots,Y_{A+n-1}$ when no change occurs, that is a $P_0$-product distribution. 

By definition of achievable sampling rates $\{\rho_n\}$ we have
\begin{align}\label{lonz2}
1-o(1)&\leq \pr(|{\cal{S}}^{\tau_n}|\leq \tau_n \f_n).
\end{align}
The following lemma, proved thereafter, says if \eqref{rdel2} holds then with probability tending to one the detector samples only $P_0$-distributed samples with probability tending to one:
\begin{lemma}\label{samlemma}
For any $\alpha>0$, if $\rho_n=o(1/n)$ then 
\begin{align}\label{ngh2}
\pr(\{\nu_n,\nu_n+1, \ldots, \nu_n+ n-1 \}&\cap {\cal{S}}^{\tau_n}=\emptyset)\notag \\
&\geq 1-o(1).
\end{align}
\end{lemma}

 This, as we now show, implies that the delay is exponential.

On the one hand, since the probability of false-alarm vanishes, we have
\begin{align*}
o(1)&\geq \pr(\tau_n<\nu_n) \nonumber \\
&\geq \pr(\tau_n<A_n/2|\nu_n\geq  A_n/2 )/2\nonumber \\
&=\pr_0(\tau_n< A_n/2)/2.
\end{align*}
This implies  
$$\pr_0(\tau_n<  A_n/2)\leq o(1),$$
and, therefore,
\begin{align}\label{faal}
\pr(\tau_n\geq   A_n/2)&\geq \pr(\tau_n\geq  A_n/2| \nu_n >  A_n/2)/2\nonumber \\
&=\pr_0(\tau_n\geq  A_n/2)/2\nonumber \\
&=1/2 -o(1).
\end{align}
Now, define events
\begin{itemize}
\item[]
 ${\cal{A}}_1\defeq\{\tau_n\geq   A_n/2\}$,
 \item[]
  ${\cal{A}}_2\defeq \{ |{\cal{S}}^{\tau_n}|\leq \tau_n \f_n\}$,
  \item[] 
  ${\cal{A}}_3\defeq \{\{\nu_n,\nu_n+1, \ldots, \nu_n+ n-1 \}\cap {\cal{S}}^{\tau_n}=\emptyset\}$,
\end{itemize}
and let
$  {\cal{A}} \defeq {\cal{A}}_1\cap {\cal{A}}_2\cap {\cal{A}}_3$.

From \eqref{lonz2}, \eqref{ngh2}, and \eqref{faal},  we get 
\begin{align}\label{bfromzero}
 \pr({\cal{A}})=1/2 -o(1).
 \end{align}
We now argue that when event ${\cal{A}}$ happens, the detector misses the change which might have occurred, say, before time $ A_n/4$, thereby implying a delay $\Theta(A_n)$ since $\tau_n\geq A_n/2$ on~${\cal{A}}$.
 
When event ${\cal{A}}$ happens, the detector takes $o(A_n/n)$ samples (this follows from event ${\cal{A}}_2$ since by assumption $\f_n=o(1/n)$). Therefore, within $\{1,2,\ldots,A_n/4\}$ there are at least $ A_n/4-o(A_n))$ time intervals of length $n$ that are unsampled. Each of these corresponds to a possible change. 
 Therefore, conditioned on event ${\cal{A}}$, with probability at least $1/4-o(1)$ the change happens before time $ A_n/4$, whereas $\tau_n \geq  A_n/2$. Hence the delay is $\Theta(A_n)$, since the probability of ${\cal{A}}$ is asymptotically bounded away from zero by \eqref{bfromzero}. \hfill $\small\blacksquare$

\begin{IEEEproof}[Proof of Lemma~\ref{samlemma}]
We have
\begin{align}\label{zztop2}
&\pr(\{\nu_n,\nu_n+1, \ldots, \nu_n+ n-1 \}\cap {\cal{S}}^{\tau_n}=\emptyset)\nonumber \\
&= \pr(\{\{\nu_n,\nu_n+1, \ldots, \nu_n+ n-1 \}\cap {\cal{S}}^{\nu_n+n-1}=\emptyset\})\nonumber \\
&\geq \pr(\{\{\nu_n,\nu_n+1, \ldots, \nu_n+ n-1 \}\cap {\cal{S}}^{\nu+n-1}=\emptyset\}\notag \\
&\hspace{3cm}\cap\{ |{\cal{S}}^{\nu_n+n-1}|\leq k\})\nonumber \\
&{=}\sum_{s: |s|\leq k}\sum_{j\in {\cal{J}}_{s}}\pr({\cal{S}}^{\nu_n+n-1}=s,\nu_n=j)\nonumber \\
&=\sum_{s: |s|\leq k}\sum_{j\in {\cal{J}}_{s}} \pr_0( {\cal{S}}^{\nu_n+n-1}=s) \pr(\nu_n=j)\nonumber \\
&\geq \frac{A_n-k\cdot n}{A_n}\sum_{s: |s|\leq k} \pr_0( {\cal{S}}^{\nu_n+n-1}=s) \nonumber \\
&=\frac{A_n-k\cdot n}{A_n} \pr_0(|{\cal{S}}^{\nu_n+n-1}|\leq  k)
\end{align}
for any $k \in \{1,2,\ldots,A_n\}$, where we defined the set of indices 
$${\cal{J}}_{s}\defeq \{ j: \{j,j+1,\ldots, j+n-1\}\cap s=\emptyset\}\}.$$
The first equality in \eqref{zztop2} holds by the definition of ${\cal{S}}^t$ (see \eqref{st}) and by \eqref{trunsam2}. The third equality  holds because event $\{{\cal{S}}^{\nu+n-1}=s\}$ involves random variables whose indices are not in $ {{\cal{J}}_{s}}$. Hence samples in $s$ are all distributed according to the nominal distribution $\pr_0$ ($P_0$-product distribution). The last inequality holds by the property \begin{align}\label{props}|{\cal{S}}^{a+b}|\leq|{\cal{S}}^{a}|+b
\end{align}
which follows from the definition of  ${\cal{S}}^{t}$.

Since $\tau_n\leq A_n+n-1$ from \eqref{lonz2} we get
\begin{align}\label{lonz3}
1-o(1)&\leq  \pr(|{\cal{S}}^{\tau_n}|\leq (A_n+n-1) \f_n)\nonumber \\
&\leq  \pr(|{\cal{S}}^{\nu_n-1}|\leq (A_n+n-1) \f_n)
\end{align}
where the second inequality holds by \eqref{trunsam2}.

Now,
\begin{align}
 \pr&(|{\cal{S}}^{\nu_n-1}|\leq (A_n+n-1) \f_n)\notag \\
& =\sum_{t=1}^{A_n}\pr(|{\cal{S}}^{t-1}|\leq (A_n+n-1) \f_n,\nu_n=t)\notag \\
& =\sum_{t=1}^{A_n}\pr_0(|{\cal{S}}^{t-1}|\leq (A_n+n-1) \f_n)\pr(\nu_n=t)\notag \\
& \leq \sum_{t=n}^{A_n+n-1}\pr_0(|{\cal{S}}^{t-1}|\leq (A_n+n-1) \f_n)\pr(\nu_n=t)\notag \\
&+\sum_{t=1}^{n-1}\pr(\nu_n=t)\notag \\
& \leq \pr_0(|{\cal{S}}^{\nu_n+n-1}|\leq (A_n+n-1) \f_n)\notag \\
&+n/A_n \notag
\\
& \leq \pr_0(|{\cal{S}}^{\nu_n+n-1}|\leq (A_n+n-1) \f_n)\notag \\
&+o(1)
\label{jsk}
\end{align}
where the last equality holds since $A_n=2^{\alpha n}$.

From \eqref{jsk} and \eqref{lonz3} we have
\begin{align}\label{cqbes}
1-o(1)\leq \pr_0(|{\cal{S}}^{\nu_n+n-1}|\leq (A_n+n-1) \f_n).
\end{align}
Letting\begin{align}
k&\defeq k_n\defeq (A_n+n-1) \f_n ,
\end{align}
and assuming that $\rho_n=o(1/n)$ we get
$$k_n\cdot n =o(A_n)$$
and hence
from \eqref{zztop2} and \eqref{cqbes}
\begin{align*}
\pr(\{\nu_n,\nu_n+1, \ldots, \nu_n+ n-1 \}&\cap {\cal{S}}^{\tau_n}=\emptyset)\notag \\
&\geq 1-o(1)
\end{align*}
which concludes the proof.
\end{IEEEproof}

\subsection{Proof of Theorem~\ref{unowaa}: Achievability}
\label{ilcuore}

We describe a detection procedure that asymptotically achieves minimum delay $n^*(\alpha)$ and any sampling rate that is $\omega(1/n)$ whenever $\alpha \in (0,D(P_0||P_1))$.

Fix $\alpha\in (0,D(P_1||P_0))$ and pick $\varepsilon>0$ small enough so that
 \begin{align}\label{eps10}
n^*(\alpha)(1+2\varepsilon)\leq n.
\end{align} Suppose we want to achieve some sampling rate $\f_n=f(n)/n$ where $f(n)=\omega(1)$ is some arbitrary increasing function (upper bounded by $n$ without loss of generality). For concreteness, it might be helpful for the reader to take $f(n)=\log \log \log \log (n)$.
 Define
$$\bar{\Delta}(n)\defeq n/f(n)^{1/3}$$
 $$\text{$s$-instants}\defeq\{t=j\bar{\Delta}(n), j\in {\mathbb{N}^*} \},$$
and recursively define
$$\Delta_0(n)\defeq f(n)^{1/3}$$
$$\Delta_i(n)\defeq \min\{2^{c\Delta_{i-1}(n)},n^*(\alpha)(1+\varepsilon)\}$$
for $i\in {1,2,\ldots, \ell}$ where $\ell$ denotes the smallest integer such that $\Delta_{\ell}(n)=n^*(\alpha)(1+\varepsilon)$. The constant $c$ in the definition of $\Delta_i(n)$ can be any fixed value such that $$0<c < D(P_1||P_0).$$

The detector starts sampling in phases at the first $s$-instant (\ie, at time $t=\bar{\Delta}(n)$) as follows:

\begin{itemize}
\item[{$1$}] {\textbf{  Preamble detection (phase zero)}}: Take  $ \Delta_0(n)$ consecutive samples and check if they are typical with respect to $P_1$. 
If the test is negative, meaning that $ \Delta_0(n)$ samples are not typical,  skip samples until the next $s$-instant and repeat the procedure \ie   sample and test $\Delta_0(n)$ observations. If the test is positive, proceed to confirmation phases. 
\item[{$2$}] {\textbf{Preamble confirmations (variable duration, $\ell-1$ phases at most)}}:
Take another $\Delta_{1}(n)$ consecutive samples and check if they are typical with respect to $P_1$. If the test is negative, skip samples until the next $s$-instant and repeat Phase zero (that is, test $\Delta_0(n)$ samples). If the test is positive, perform a second confirmation phase with $\Delta_{1}(n)$ replaced with $\Delta_{2}(n)$, and so forth. Note that each confirmation phase is performed on a new set of samples.  If $\ell-1$ consecutive confirmation phases (with respect to the same $s$-instant) are positive, the receiver moves to the full block sampling phase. 
  \item[{$3$}] {\textbf{Full block sampling ($\ell$-th phase): }} Take another $$\Delta_{\ell}(n)=n^*(\alpha)(1+\varepsilon)$$ samples and check if they are typical with respect to $P_1$. If they are typical, stop. Otherwise, skip samples until the next $s$-instant and repeat Phase zero. If by time $A_n+n-1$ no sequence is found to be typical, stop.
 \end{itemize}
Note that with our  $f(n)=\log \log \log \log (n)$ example, we have two preamble confirmation phases followed by the last full block sampling phase.

For the probability of false-alarm we have 
\begin{align}\label{falsoal}
\pr(\tau_n<\nu_n) &\leq  2^{\alpha n}\cdot
2^{- n^*(\alpha)(1+\varepsilon)(D(P_1||P_0)-o(1))}\nonumber \\
&=2^{-n\alpha\Theta(\varepsilon)}\nonumber \\
&=o(1)
\end{align}
because whenever the detector stops, the previous $$n^*(\alpha)(1+\varepsilon)$$ samples are necessarily typical with respect to $P_1$.  Therefore, the inequality \eqref{falsoal} follows from \eqref{mismatch} and a union bound over time indices. The equality in \eqref{falsoal} follows directly from the definition of $n^*(\alpha)$ (see \eqref{nstaar}).

Next, we analyze the delay of the proposed scheme. We show that
\begin{align}\label{delare}
\pr(\tau_n\leq \nu_n+(1+2\varepsilon)n^*(\alpha))=1-o(1).
\end{align}
To see this, note that by the definition of $\bar{\Delta}(n)$ and because each $\Delta_i(n)$ is exponentially larger than the previous $\Delta_{i-1}(n)$, 
$$\bar{\Delta}(n)+\sum_{i=0}^\ell \Delta_i(n)\leq (1+2\varepsilon)n^*(\alpha) $$ for $n$ large enough.
Applying \eqref{typset2} and taking a union bound, we see that when the samples are distributed according to $P_1$, the series of $\ell+1$ hypothesis tests will all be positive with probability $1-o(1)$. Specifically,
\begin{align}\label{delare2}
\pr(\mbox{any test fails}) \leq \sum_{i=0}^{\ell} O\left(\Delta_i(n)\right)^{-\frac{1}{3}} = o(1).
\end{align}
 Since $\varepsilon$ can be made arbitrarily small, from \eqref{falsoal} and \eqref{delare} we deduce that the detector achieves minimum delay (see Theorem~\ref{unow}, Claim 2)) .

Finally, to show that the above detection procedure achieves sampling rate $$\f_n=f(n)/n$$ we need to establish that \begin{align}\label{Z2}
\pr(|\cS^{\tau_n}|/\tau_n\geq \f_n)\overset{n\to \infty}{\longrightarrow} 0.
\end{align}
To prove this, we first compute the sampling rate of the detector when run over an i.i.d.-$P_0$ sequence, that is, a sequence with no transient change. As should be intuitively clear, this will essentially give us the desired result, since in the true model, the duration of the transient change, $n$, is negligible with respect to $A_n$ anyway.

To get a handle on the sampling rate of the detector over an i.i.d.-$P_0$ sequence, we start by computing the expected number of samples $N$ taken by the detector at any given $s$-instant, when the detector is started at that specific $s$-instant and the observations are all i.i.d. $P_0$. Clearly, this expectation does not depend on the $s$-instant.\footnote{Boundary effects due to the fact that $A_n$ need not be a multiple of $\bar{\Delta}_n$ play no role asymptotically and thus are ignored.} 
We have 
\begin{align}\label{master11}
\ex_0 N \leq \Delta_0(n)+\sum_{i=0}^{\ell-1} p_{i} \cdot \Delta_{i+1}(n)
\end{align} where $p_i$ denotes the probability that the $i$-th confirmation phase is positive given that the detector actually reaches the  $i$-th confirmation phase, and $\ex_0$ denotes expectation with respect to an i.i.d.-$P_0$ sequence. Since each phase uses new, and therefore, independent, observations, from \eqref{mismatch} we conclude that 
$$p_i\leq 2^{-\Delta_{i}(n)(D(P_1||P_0)-o(1))}.$$
Using the definition of $\Delta_{i}(n)$, and recalling that $0 < c < D(P_1||P_0)$, this implies that the sum
in the second term of \eqref{master11} is negligible, and
\begin{align}\label{master22}
\ex_0 N_s=\Delta_0(n)(1+o(1)).\end{align}
Therefore, the expected number of samples taken by the detector up to any given time $t$ can be upper bounded as
\begin{align}\label{fnn}
\ex_0 |\cS^{t}|&\leq \frac{t}{\bar{\Delta}(n)}\Delta_0(n)(1+o(1))\nonumber \\
&=t \frac{f(n)^{2/3}}{n}(1+o(1)).
\end{align}
This, as we now show, implies that the detector has the desired sampling rate. 
We have
\begin{align}\label{sramdet2}
\pr&(|{{\cS}}^{\tau_n}|/\tau_n\geq \f_n)\notag\\
&\leq \pr(|{{\cS}}^{\tau_n}|/{\tau}_n\geq \f_n, \nu_n \leq \tau_n \leq
\nu_n+(1+2\varepsilon)n^*(\alpha))\notag\\
&+1-\pr(\nu_n \leq \tau_n \leq \nu_n+ (1+2\varepsilon)n^*(\alpha))\notag\\
&\leq \pr(|{{\cS}}^{\tau_n}|/\tau_n\geq \f_n, \nu_n \leq \tau_n \leq
\nu_n+n)\notag\\
&+1-\pr(\nu_n \leq \tau_n \leq \nu_n+ (1+2\varepsilon)n^*(\alpha))
\end{align}
where the second inequality holds for $\varepsilon$ small enough by the definition of $n^*(\alpha)$.

The fact that \begin{align}\label{igw2}
1-\pr(\nu_n\leq \tau_n<\nu_n+ (1+2\varepsilon)n^*(\alpha))=o(1)
\end{align}
follows from \eqref{falsoal} and \eqref{delare}. For the first term on the right-hand side of the second inequality in \eqref{sramdet2}, we have
\begin{align}\label{dex2}
\pr&(|{{\cS}}^{\tau_n}|/\tau_n\geq \f_n, \nu_n \leq \tau_n \leq
\nu_n+n)\nonumber \\
&\leq \pr(|\cS^{\nu_n+n}|\geq \nu_n\f_n)\notag\\
&\leq \pr(|\cS^{\nu_n-1}|\geq \nu_n\f_n-n-1).
\end{align}
Since $\cS_{\nu_n-1}$ represents sampling times before the transient change, the underlying process is  i.i.d. $P_0$, so we can use our previous bound on the sampling rate to analyze $\cS_{\nu_n-1}$. Conditioned on 
reasonably large values of $\nu_n$, in particular, all $\nu_n$ satisfying \begin{align}\label{hypothesa}\nu_n\geq \sqrt{A_n=2^{\alpha n}}\end{align}
 we have 
\begin{align}\label{primamon2}
\pr(|\cS^{ \nu_n-1}|\geq \nu_n&\f_n-n-1|\nu_n)\leq \frac{\ex_0|\cS^{ \nu_n}|} {\nu_n\f_n-n-1 }\nonumber \\
&\leq  \frac{{f(n)^{2/3}}(1+o(1)) }{n(\f_n-(n+1)/\nu_n)}\nonumber \\
&\leq  \frac{{f(n)^{2/3}}(1+o(1)) }{n\f_n(1-o(1))}\nonumber \\
&=  \frac{(1+o(1)) }{f(n)^{1/3}(1-o(1))}\nonumber \\
&=o(1)
\end{align}
where the second inequality holds by \eqref{fnn}; where the third inequality holds by \eqref{hypothesa} and because $\f_n=\omega(1/n)$; and where the last two equalities hold by the definitions of $\f_n$ and $f(n)$. 

Removing the conditioning on $\nu_n$,
\begin{align}\label{barito2}\pr&(|\cS^{\nu_n-1}|\geq \nu_n\f_n-n-1)\nonumber \\
&\leq 
\pr(|\cS^{\nu_n-1}|\geq \nu_n\f_n-n-1,\nu_n\geq \sqrt{A_n})\notag\\
&+\pr(\nu_n< \sqrt{A_n})\notag \\
&=o(1)
\end{align}
by \eqref{primamon2} and the fact that $\nu_n$ is uniformly distributed over $\{1,2,\ldots,A_n\}$. Hence, from \eqref{dex2}, the first term on the right-hand side of the second inequality in \eqref{sramdet2} vanishes. 

This yields \eqref{Z2}.

\subsection{Discussion}\label{disc}
There is obviously a lot of flexibility around the quickest detection procedure described in Section~\ref{ilcuore}. Its main feature is the sequence of binary hypothesis tests, which manages to reject the hypothesis that a change occurred with as few samples as possible when the samples are drawn from $P_0$, while maintaining a high probability of detecting the transient change.

 It may be tempting to simplify the detection procedure by considering, say, only two phases, a preamble phase and the full block phase. Such a scheme, which is similar in spirit to the one proposed in \cite{6960771}, would not work, as it would produce either a much higher level of false-alarm, or a much higher
sampling rate. We provide an intuitive justification for this below, thereby highlighting the role of the multiphase procedure.
 
 Consider a two phase procedure, a preamble phase followed by a full block phase. Each time we switch to the second phase, we take $\Theta(n)$ samples. Therefore, if we want to achieve a vanishing sampling rate, then necessarily the probability of switching from the preamble phase to the full block phase under $P_0$  should be $o(1/n)$. By Sanov's theorem, such a probability can be achieved only if the preamble phase makes it decision to switch to the full block phase based on at least $\omega(\log n)$ samples, taken over time windows of size $\Theta(n)$.
This translates into a sampling rate of $\omega((\log n)/n)$ at best, and we know that this is suboptimal, since any sampling rate $\omega(1/n)$ is achievable.  

The reason a two-phase scheme does not yield a sampling rate lower than $\omega((\log n)/n)$ is that it is too coarse. To guarantee a vanishing sampling rate, the decision to switch to the full block phase should be based on at least $\log(n)$ samples, which in turn yields a suboptimal sampling rate.  
The important observation is that the (average) sampling rate of the two-phase procedure essentially corresponds to the sampling rate of the first phase, but the first phase also controls the decision to switch to the full block phase and sample continuously for a long period of order $n$. In the multiphase procedure, however, we can separate these two functions. The first phase controls the sampling rate, but passing the first phase only leads us to a second phase, a much less costly decision than immediately switching to full block sampling. By allowing multiple phases, we can ensure that when the decision to ultimately switch to full sampling occurs, it only occurs because we have accumulated a significant amount of evidence that we are in the middle of the transient change. In particular, note that many other choices would work for the length and probability thresholds used in each phase of our sampling scheme. The main property we rely on is that the lengths and probability thresholds be chosen so that the sampling rate is dominated by the first phase.



\subsection{Proof of Theorem~\ref{thm}}
\label{app1}
In this section, we prove Theorem~\ref{thm}. A reader familiar with the proofs 
presented in \cite{6960771} will recognize Theorem~\ref{thm} as a corollary of 
Theorem~\ref{unowaa}, but we include a detailed proof below for interested 
readers unfamiliar with the prior work \cite{6960771}.

\subsubsection{Converse of Theorem~\ref{thm}}
By using the same arguments as for Lemma~\ref{samlemma}, and simply replacing replacing $n$ with $d_B$, one readily sees that if \begin{align}\label{rdel}
\rho_Bd_B=o(1)
\end{align} 
then
\begin{align}\label{zztop22}
\pr(\{\nu_B,&\nu_B+1, \ldots, \nu_B+ d_B-1 \}\cap {\cal{S}}^{\tau_B}=\emptyset)\nonumber \\
&\geq (1-o(1)).
\end{align}
Since the decoder samples no codeword symbol with probability approaching one, the decoding error probability will tend to one whenever the rate is positive (so that $(M-1)/M$ tends to one).  

\subsubsection{Achievability of Theorem~\ref{thm}}

Fix $\beta>0$.  We show that any 
$\bm{R}\in (0,\bm{C}(\beta)]$ is achievable with codes $\{\cC_B\}$ whose delays satisfy $d(\cC_B,\varepsilon_B)\leq n^*_B(\beta,\bm{R})(1+o(1))$ whenever the sampling rate $\f_B$ is such that $$\f_B=\frac{f(B)}{B}$$ for some $f(B)=\omega(1)$.

Let $X\sim P$ be some channel input and let $Y$ denote the corresponding output, \ie $(X,Y)\sim P(\cdot)Q(\cdot|\cdot)$. For the moment we only assume that $X$ is such that $I(X;Y)>0$. Further, we suppose that the codeword length $n$ is linearly related to $B$, \ie $$\frac{B}{n}=q $$ for some fixed constant $q>0$. We shall specify this linear dependency later to accommodate the desired rate $\bm{R}$. Further, let $$\tilde{f}(n)\defeq  f(q\cdot n)/q$$
and  $$\tilde{\f}_n\defeq \frac{\tilde{f}(n)}{n}.$$
Hence, by definition we have
$$\tilde{\f}_n=\f_B.$$

Let $a$ be some arbitrary fixed input symbol such that $$Q(\cdot|a)\ne Q(\cdot|\star).$$ Below we introduce the quantities $\bar{\Delta}(n)$ and $\Delta_{i}(n)$, $1\leq i\leq \ell$, which are defined as in Section~\ref{ilcuore} but with $P_0$ replaced with $Q(\cdot|\star)$, $P_1$ replaced with $Q(\cdot|a)$, $f(n)$ replaced with $\tilde{f}(n)$, and $n^*(\alpha)$ replaced with $n$.

\textbf{Codewords: preamble followed by constant composition information symbols.}
Each codeword $c^n(m)$ starts with a common preamble that consists of
$\bar{\Delta}(n)$ repetitions of symbol $a$. The remaining $$n-\bar{\Delta}(n)$$ components $$c_{\bar{\Delta}(n)+1}^n(m)$$ of $c^n(m)$ of each message $m$ carry information and are generated as follows.
For message $1$, randomly generate
length $n-\bar{\Delta}(n)$ sequences $x^{n-\bar{\Delta}(n)}$ i.i.d. according to $P$ until when 
$x^{n-\bar{\Delta}(n)}$ is typical with respect to $P$. In this case we let $$c_{\bar{\Delta}(n)+1}^n(1) \defeq x^{n-\bar{\Delta}(n)}\,,$$ move to message $2$, and repeat the procedure until when
a codeword has been assigned to each message. 

From \eqref{typset2}, 
for any fixed $m$ no repetition will be required to generate $c_{\bar{\Delta}(n)+1}^n(m)$ with probability tending to one as $n~\to~\infty$. Moreover, by construction the codewords are essentially of constant composition, \ie each symbol appears roughly the same number of times in all codewords, and all codewords have cost $$n\ex[k(X)](1+o(1))$$ as $n\to \infty$.

\textbf{Codeword transmission time.}
Define the set of start instants  
$$\text{$s$-instants}\defeq\{t=j\bar{\Delta}(n), j\in {\mathbb{N}}^* \}.$$
Codeword transmission start time $\sigma(m,\nu_n)$ corresponds to the first $s$-instant $\geq \nu_n$ (regardless of $m$).

\textbf{Sampling and decoding procedures.}
The decoder first tries to detect the preamble by using a similar detection procedure as in the achievability of Theorem~\ref{unowaa}, then applies a standard message decoding isolation map. 

 Starting at the first $s$-instant (\ie at time $t=\bar{\Delta}(n)$), the decoder samples in phases as follows.
\begin{itemize}
\item[{$1$}] {\textbf{  Preamble test  (phase zero)}}: Take  $ \Delta_0(n)$ consecutive samples and check if they are typical with respect to $Q(\cdot|a)$. 
If the test turns negative, the decoder skips samples until the next $s$-instant when it repeats the procedure. If the test turns positive, the decoder moves to the confirmation phases. 
\item[{$2$}] {\textbf{Preamble confirmations (variable duration, $\ell-1$ phases at most)}}:
The decoder takes another $\Delta_{1}(n)$ consecutive samples and checks if they are typical with respect to $Q(\cdot|a)$. If the test turns negative the decoder skips samples until the next $s$-instant when it repeats Phase~zero (and tests $\Delta_0(n)$ samples). If the test turns positive, the decoder performs a second confirmation phase based on new $\Delta_{2}(n)$ samples, and so forth.  If $\ell-1$ consecutive confirmation phases (with respect to the same $s$-instant) turn positive, the decoder moves to the message sampling phase. 
  \item[{$3$}] {\textbf{Message sampling and isolation ($\ell$-th phase): }} Take another $n$ samples and check if among these samples there are $n-\bar{\Delta}(n)$ consecutive samples that are jointly typical with the $n-\bar{\Delta}(n)$ information symbols of  one of the codewords. If one codeword is typical, stop and declare the corresponding message. If more than one codeword is typical declare one message at random. If no codeword is typical,  the decoder stops sampling until the next $s$-instant and repeats Phase zero. If by time $A_B+n-1$ no codeword is found to be typical, the decoder declares a random message.
 \end{itemize}

\textbf{Error probability.}
Error probability and delay are evaluated in the limit $B\to \infty$ with  $A_B = 2^{\beta B}$ and with 
 \begin{align}
\label{cn1}
q=\frac{B}{n} <\min\bigg\{I(X;Y),
\frac{I(X;Y)+D(Y||{Y_\star})}{1+\beta}\bigg\}.
\end{align}

We first compute the error probability averaged over codebooks and messages.  
Suppose message $m$ is  transmitted and denote by $\E_m$ the error event that the decoder stops and outputs a message $m'\ne m$. Then we have
\begin{align}\label{decom}
\E_m\subseteq\E_{0,m}\cup_{m'\ne m}(\E_{1,m'}\cup\E_{2,m'}),
\end{align}
where events $\E_{0,m}$,  $\E_{1,m'}$, and $\E_{2,m'}$ are defined as
\begin{itemize}
\item $\E_{0,m}$: at the $s$-instant corresponding to $\sigma$, the preamble test phase or one of the preamble confirmation phases turns negative, or $c^{n}_{\bar{\Delta}(n)+1}(m)$ is not found to be typical by time $\sigma+n-1$;
\item $\E_{1,m'}$: the decoder stops at a time $t<\sigma$ and declares $m'$;
\item $\E_{2,m'}$: the decoder stops at a time $t$ between $\sigma$ and $\sigma + n-1$ (including $\sigma$ and $\sigma+n-1$) and declares~$m'$.
\end{itemize}

From Sanov's theorem, 
\begin{align}\label{eroo}
\pr_m(\E_{0,m})= \varepsilon_1(B)
\end{align}
where $\varepsilon_1(B)=o(1)$. Note that this equality holds pointwise (and not only on average over codebooks) for any 
specific (non-random) codeword $c^n(m)$ since, by construction, they all satisfy the constant composition property
\begin{align}\label{costoco}||\hat{P}_{c^n_{\bar{\Delta}+1}(m)}-P||\leq (n-\bar{\Delta})^{-1/3}=o(1)\end{align}
as $n\to \infty$.

Using analogous arguments as in the achievability of \cite[Proof of Theorem 1]{6397617}, we obtain the upper bounds
$$\pr_m(\E_{1,m'}) \leq  2^{\beta B}\cdot
2^{- n(I(X;Y)+D(Y||Y_\star)-o(1))}$$
and
$$\pr_m(\E_{2,m'}) \leq  2^{-n(I(X;Y)-o(1))}$$
which are both valid for any fixed $\varepsilon>0$ provided that $B$ is large enough. Hence from the union bound
\begin{align*}\pr_m(\E_{1,m'}\cup\E_{2,m'}) \leq &2^{- n(I(X;Y)-o(1))} \\
&+ 2^{\beta B} \cdot 2^{-n(I(X;Y) +
D(Y||{Y_\star})-o(1))}\,.\end{align*}
Taking a second union bound over all possible wrong
messages, we get
\begin{align}\label{er11}\pr_m( \cup_{m'\ne m}(\E_{1,m'}&\cup\E_{2,m'}))\leq 2^B\Big(2^{- n(I(X;Y)-o(1))}
\nonumber\\
+  &2^{\beta B} \cdot 2^{-n(I(X;Y) +
D(Y||{Y_\star})-o(1))}\Big)\nonumber\\
&\defeq \varepsilon_2(B)
\end{align}
where $ \varepsilon_2(B)=o(1)$ because of \eqref{cn1}.  

Combining \eqref{decom}, \eqref{eroo}, \eqref{er11}, we get from the union bound
\begin{align}\label{er33}
\pr_m(\E_m)&\leq \varepsilon_1(B)+\varepsilon_2(B)\nonumber \\
&=o(1)
\end{align}
for any $m$.

\textbf{Delay.}
We now show that  the delay of our coding scheme is at most $n(1+o(1))$. Suppose 
codeword $c^n(m)$ is sent.
If $$\tau_B> \sigma + n$$ then necessarily $c^n_{\bar{\Delta}+1}(m)$ is not typical with the corresponding channel outputs. Hence
\begin{align}
\pr_m(\tau_B-\sigma \leq n)&\geq 1-\pr_m(\E_{0,m})\nonumber \\
&=1-\varepsilon_1(B)\label{dela}
\end{align}
 by \eqref{eroo}.
Since $ \sigma \leq \nu_B+\bar{\Delta}(n)$ and $\bar{\Delta}(n)=o(n)$ we get\footnote{Recall that $B/n$ is kept fixed and $B\to \infty$.}
$$\pr_m(\tau_B-\nu_B \leq n(1+o(1)))\geq 1-\varepsilon_1(B)\,.$$
Since this inequality holds for any codeword $c^n(m)$ that satisfies \eqref{costoco}, the delay is no more than $n(1+o(1))$.
 Furthermore, from \eqref{er33} there exists a specific non-random code $\cC$ whose error
probability, averaged over messages, is less than
$\varepsilon_1(n)+\varepsilon_2(n)=o(1)$ whenever condition \eqref{cn1} is satisfied.  Removing the half of the codewords with the highest error
probability, we end up with a set ${\cal{C}}'$ of $2^{B-1}$ codewords 
whose maximum error probability satisfies
\begin{align}\label{mwi}\max_m\pr_m(\E_m)\leq
o(1)
\end{align}
 whenever condition \eqref{cn1} is satisfied.

Since any codeword has cost $n\ex[k(X)](1+o(1))$, condition \eqref{cn1} is equivalent to
 \begin{align}\label{eqmaster}
\bm{R}<\min&\bigg\{\frac{I(X;Y)}{\ex[\ck(X)](1+o(1))},\notag \\
&\frac{I(X;Y)+D(Y||{Y_\star})}{\ex[\ck(X)](1+o(1))(1+\beta)}\bigg\}
\end{align}
where $$\bm{R}\defeq \frac{B}{\cK({\cal C}')}$$
denotes the rate per unit cost of ${\cal C}'$.

Thus, to achieve a given $\bm{R}\in (0,\bm{C}(\beta))$ it suffices to choose the input distribution and the codeword length as $$X=\arg\max \{ \ex [k(X')]: {X'}\in {\cal{P}}(\bm{R})\}$$ 
and $$n=n_B^*(\beta,\bm{R})$$
(see \eqref{ennstar} and \eqref{probset}). By a previous argument the corresponding delay is no larger than $n_B^*(\beta,\bm{R})(1+o(1))$.

\textbf{Sampling rate.}
For the sampling rate, a very similar analysis to the achievability proof of Theorem~\ref{unowaa} (see from equation \eqref{Z2} onwards with $f(n)$, $\f_n$, $n^*(\alpha)$, and $A_n$ replaced with $\tilde{f}(n)$, $\tilde{\f}_n$, $n^*(\beta,\bm{R})$, and $A_B$, respectively)
shows that
\begin{align}\label{Z}
\pr_{m}(|\cS^{\tau_B}|/\tau_B\geq \f_B)\overset{B\to \infty}{\longrightarrow} 0.
\end{align}

Note that the arguments that establish \eqref{Z} rely only on the preamble detection procedure. In particular, they do not use \eqref{eqmaster} and hold for any codeword length $n_B$ as long as $n_B= \Theta(B)$.

\section{Conclusion}
We have proved an essentially tight characterization of the sampling rate 
required to have no capacity or delay penalty for the asynchronous 
communication model of \cite{6960771}. The key ingredient in our results is
a new, multi-phase, adaptive sampling scheme used to detect when the received
signal's distribution switches from the pure noise distribution to the
codeword distribution. As noted above, there is a lot of flexibility around the quickest detection procedure described in Section~\ref{ilcuore}, but a simple, two level generalization of the sampling algorithm from \cite{6960771} is insufficient to achieve the optimal sampling rate. Instead, a fine-grained, multi-level scheme is needed.

\bibliographystyle{plain}

\bibliography{../../../common_files/bibiog}

\end{document}